\documentclass[useAMS,usenatbib]{mn2e}

\usepackage{amsmath,amssymb,wasysym,psfig,graphicx,lscape,natbib,epsfig}
\usepackage[totalwidth=480pt, totalheight=680pt]{geometry}
\bibpunct{(}{)}{;}{a}{}{,}

\def\mn={MNRAS}

\title[How common are long Gamma-Ray Bursts in the Local Universe?]{How common are long Gamma-Ray Bursts in the Local Universe?}
\author[Chapman R. and others]{Robert Chapman$^1$\thanks{Email:r.1.chapman@herts.ac.uk}, Nial R. Tanvir$^2$, Robert S. Priddey$^1$ and Andrew J. Levan$^3$\\
$^1$Centre for Astrophysics Research, University of Hertfordshire, College Lane, Hatfield AL10~9AB, UK\\
$^2$Department of Physics and Astronomy, University of Leicester, Leicester, LE1~7RH, UK\\
$^3$Department of Physics, University of Warwick, Coventry, CV4 7AL, UK}

\begin{document}

\date{20th June 2007}

%\pagerange{\pageref{firstpage}--\pageref{lastpage}} \pubyear{2002}

\maketitle

\label{firstpage}

\begin{abstract}
The two closest Gamma-Ray Bursts so far detected (GRBs\,980425 \& 060218) were both under-luminous, spectrally soft, long duration bursts with smooth, single-peaked light curves. Only of the order of 100 GRBs have measured redshifts, and there are, for example, 2704 GRBs in the BATSE catalogue alone. It is therefore plausible that other nearby GRBs have been observed but not identified as relatively nearby. Here we search for statistical correlations between BATSE long duration GRBs and galaxy samples with recession velocities $v\le 11,000\rm{km~s}^{-1}$~($z=0.0367, \approx155~\rm{Mpc}$) selected from two catalogues of nearby galaxies. We also examine the correlations using burst sub-samples restricted to those with properties similar to the two known nearby bursts. Our results show correlation of the entire long GRB sample to remain consistent with zero out to the highest radii considered whereas a sub-sample selected to be low fluence, spectrally soft, with smooth single-peaked light curves (177 bursts) demonstrates increased correlation with galaxies within $\approx155~\rm{Mpc}$. The measured correlation ($28\%\pm16\%$ of the sample) suggests that BATSE observed between 2 and 9 long duration GRBs per year similar to, and from within similar distances to GRBs\,980425 and 060218. This implies an observed local rate density (to BATSE limits) of $700\pm360~\rm{Gpc}^{-3}\rm{yr}^{-1}$ within 155~Mpc.
\end{abstract}


\section{Introduction}

The two closest Gamma-Ray Bursts (GRBs) detected so far are GRB\,980425~\citep{1998Natur.395..670G} and GRB\,060218~\citep{2006GCN..4775....1C,2006GCN..4792....1M}, both long duration GRBs of exceptionally low luminosity. Indeed GRB\,980425 is usually taken as the archetypal low-luminosity, low variability, soft spectrum GRB. During nine years of operation, BATSE (the Burst and Transient Source Experiment on board the \textit{Compton Gamma-Ray Observatory}) detected 2704 GRBs. As is well known, there is good evidence for two observed classes of GRB distinguished most clearly by $\rm{T}_{90}$, the time taken to collect $90\%$ of a burst's total gamma-ray fluence~\citep{1993ApJ...413L.101K,1984Natur.308..434N}. Long duration GRBs (L-GRBs, $\rm{T}_{90}>2\rm{s}$) show a correlation between spectral peak energy and overall isotropic energy release - the $E_p/E_{iso}$ or Amati relation~ \citep{2002A&A...390...81A} - where softer bursts are less isotropically energetic (although GRB\,980425 is an outlier to this relationship in the sense of being too spectrally hard for its isotropic luminosity, it is still softer than the majority of BATSE long bursts). L-GRBs also exhibit a spectral lag-luminosity relationship~\citep{2000ApJ...534..248N,2002ApJ...569..682S,2002ApJ...579..386N} where there is anti-correlation between overall luminosity and the time delay between the observation of corresponding light curve features in different energy bands. Furthermore, GRBs with long spectral lags (and hence low luminosity) tend to have smoother light curves with broader features, and lower peak fluxes~\citep{2001AIPC..587..176N}.  Isotropic-equivalent peak luminosity has also been shown to correlate with light curve variability, originally for a relatively small number of bursts~\citep{2001ApJ...552...57R,2000astro.ph..4176F} but recently confirmed by~\citet{2005MNRAS.363..315G} for a larger sample, though with a larger scatter in the data. Though the exact form and tightness of the correlation is currently a matter of some debate~\citep{2005astro.ph..8111R,2006MNRAS.371..843G,2006MNRAS.366..219L}, there appears little doubt that the luminosity of long GRBs correlates with variability. 

GRBs\,980425, 060218 and the next nearest L-GRB with a well observed supernova component to date GRB\,031203~\citep{2003GCN..2459....1G,2004ApJ...611..200P,2004ApJ...605L.101W} are comprehensively reviewed by~\citet{2007ApJ...654..385K}. The properties of these nearby bursts, along with the correlations discussed above, suggest that under-luminous L-GRBs are likely to be spectrally soft with smooth, single-peaked light curves. Thus we can use these properties to select those bursts most likely to be of intrinsically low luminosity and therefore drawn from a relatively nearby population. Furthermore, as suggested by arguments based on analyses of detector sensitivities and luminosity function calculations~\citep{2006Natur.442.1011P, 2006Natur.442.1014S, 2007ApJ...662.1111L, 2006ApJ...645L.113C}, this sub-class of low-luminosity bursts may be many times more prevalent in the local Universe than high-luminosity bursts.

Approximately $75\%$ of the 2704 BATSE GRBs were long, but due to large localisation error regions few of these bursts have identified hosts or redshifts. However, if some originated within similar distances to GRB\,980425 and GRB\,060218 then it should be possible to estimate this fraction statistically via their distribution on the sky, using the same technique we used previously to investigate the distribution of short GRBs~\citep{2005Natur.438..991T}. If indeed there is an unidentified population of under-luminous, soft, smooth L-GRBs, then restricting the BATSE sample to those with properties similar to GRB\,980425 should enhance any correlation signal.

We therefore consider below the correlation between BATSE long bursts and two galaxy catalogues:
the PSCz galaxy redshift survey~\citep{2000MNRAS.317...55S} and the Third Reference catalogue of Bright Galaxies (\textit{RC3})~\citep{1991trcb.book.....D}. The \textit{PSCz} is based on the \textit{IRAS} Point Source catalogue, and is less affected by dust extinction in the galactic plane than other redshift surveys, making it an appropriate comparison dataset for the all-sky BATSE dataset. Though less complete, the \textit{RC3} is useful as a further comparison set and a check against any possible catalogue bias.

\section{Methods}

The long burst sample consists of the 1437 long GRBs with location errors $\le10^{\circ}$ and measured fluences in the current BATSE catalogue (4B(R)~\citet{1999ApJS..122..465P}, including web supplement). Our galaxy samples are drawn from two galaxy catalogues, the IRAS \textit{PSCz} galaxy redshift survey~\citep{2000MNRAS.317...55S} and the Third Reference Catalogue of Bright Galaxies (\textit{RC3})~\citep{1991trcb.book.....D}. Using the measured heliocentric recession velocities of galaxies in these catalogues, we made samples of galaxies within concentric spheres of recession velocity $v<2000,~5000,~8000$ and $11000~\rm{km~s^{-1}}$ (corresponding to radii of 28($z=0.0067$), 70(0.017), 113(0.027) and 155(0.037)~Mpc respectively)\footnote{Throughout this paper we use $\rm{H_0}=71\rm{km~s^{-1}~Mpc^{-1}}$. Due to individual peculiar velocities, recession velocity is not always an exact distance proxy.}. Following the method described in \citet{2005Natur.438..991T}, we define the statistic $\Phi$ for each galaxy sample compared with the long GRBs as shown in equation~\ref{phi}, where $\theta_{ij}$ is the separation between the $i$th burst and the $j$th galaxy, and $\epsilon_i$ is the error circle of the $i$th burst position (statistical combined with systematic errors from model 1 of~\citet{1999ApJS..122..503B}). For each comparison set (burst sample with galaxy sample), $\Phi$ thus provides a measure of the overall correlation between the bursts and galaxies.

\begin{equation}
\label{phi}
\Phi=\sum_{i}^{Bursts}\sum_{j}^{Galaxies}\frac{1}{\epsilon_i}\int_{\theta_{ij}}^{\infty}
\frac{1}{\sqrt{2\pi}\epsilon_i}\exp\left[\frac{-\theta^2}{2\epsilon_i^2}\right]d\theta
\end{equation}

Values of $\Phi$ can be calibrated and used to measure the percentage of bursts correlated with each galaxy sample by simulating distributions of known correlation with that galaxy sample. To achieve this we generated a number (501) of distributions of random bursts (each with the same number of positions as the number of bursts under consideration, the same positional errors, and distributed on the sky according to the known BATSE sky exposure map~\citep{2003AIPC..662..176H}). From these we computed $\Phi_0$ and its associated dispersion. Similarly, we generated 501 100$\%$ correlated test distributions for each galaxy sample (this time with ``host" positions chosen randomly from the galaxy sample and pseudo-bursts displaced from these positions assuming Gaussian probability distributions within error circles chosen randomly from the burst sample) which enabled us to calculate $\Phi_{100}$. Calibration plots of $\Phi/\Phi_0$ versus percentage correlation with associated errors could then be produced.

To reiterate, the closest known bursts (GRBs\,980425, 060218 and 031203) were under-luminous and spectrally soft with smooth single-peaked light curves. In order to select a burst sub-sample with properties similar to these bursts, three separate selections were performed on the L-GRB sample, based on observed fluence, spectral softness, and overall light curve shape. First, as argued above, since long-lag (and therefore likely under-luminous) bursts increase from a negligible proportion of BATSE bright bursts to $\sim50\%$ at trigger threshold~\citep{2002ApJ...579..386N}, we ordered the L-GRBs by total burst fluence and selected the low-fluence half. Similarly, to select for spectrally soft bursts, we split the total sample in half by the ratio of observed fluence in BATSE energy channels 1 ($20-50~\rm{keV}$) and 3 ($100-300~\rm{keV}$). We did not attempt further cuts on the data based on fluence or spectral hardness since this would not be justified for several reasons. Firstly, the number of bursts selected in finer cuts would be very small for statistical analysis. Secondly the low fluence bursts (for example) will obviously contain bursts that are dim due to being further away and these cannot be distinguished from the nearby bursts - we aimed purely to increase the fraction of nearby bursts in the samples as predicted by the correlations guiding the cuts. Thirdly, given the overall lack of correlation in the sample as a whole, any measured correlation is likely to be very weak and further cuts beyond the median would be statistically dubious. The third selection for bursts of a smooth, single-peaked nature was made by visual examination of the light curves of the entire L-GRB sample. We emphasise that this selection was based on simple pre-agreed criteria to select smooth, single peaked curves broadly similar to the known local bursts, not to attempt detailed selection with respect to small scale variation. To further minimise any subjectivity involved, selection was performed independently by two of us and then arbitrated by a third. Finally, the bursts common to all three selections (177) were then used to form a low-fluence, spectrally-soft, single-peaked sub-sample.

\section{Results}

\begin{figure}
\psfig{figure=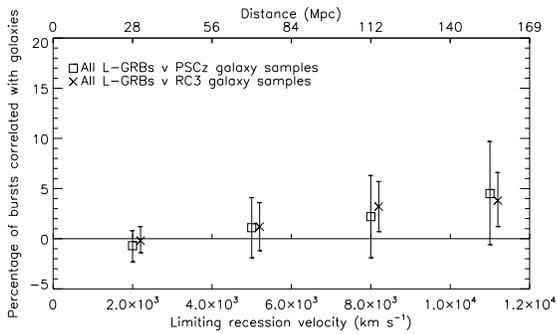, width=3.0in}
\centering{\caption[Correlation of bursts versus galaxy distance in the \textit{PSCz} and \textit{RC3} catalogues]{\label{bothcorr}Measured correlation (expressed as the percentage of bursts in each sample correlated with galaxy distribution) versus galactic recession velocity for concentric spheres of galaxy samples. Data points have been separated along the x axis for clarity, and error bars represent $1\sigma$ errors.}}
\end{figure}

Figure~\ref{bothcorr} shows plots of correlation (expressed as the percentage of bursts in each sample correlated with galaxy distribution) between BATSE long GRBs and concentric spheres of galaxy samples from both the \textit{PSCz} and \textit{RC3} catalogues. The results show a high degree of consistency, confirming that the correlation measurements are not dependent on a chance choice of galaxy catalogue. In practice, each volume sample in the \textit{RC3} catalogue contained between 1.5 and 2 times the number of galaxies in the equivalent \textit{PSCz} sample. This inevitably leads to increased dispersion in the values of $\Phi$ measured with the \textit{PSCz} samples compared to the \textit{RC3} samples, though we still consider the completeness and homogeneity of the \textit{PSCz} sample to make it a more appropriate comparison set.

As can be seen, long burst correlation (particularly with the \textit{PSCz} samples) remains formally consistent with zero out to the largest radii considered of $v\le11000~\rm{km~s^{-1}}$ ($\approx155~\rm{Mpc}$), confirming that nearby long GRBs are indeed rarely observed events. However, the most probable level of correlation increases with distance, but this could not be reliably investigated beyond the radii considered due to the flux-limited nature of the galaxy catalogues meaning that there are just too few galaxies in the catalogues at larger radii to be useful.

\begin{figure}
\psfig{figure=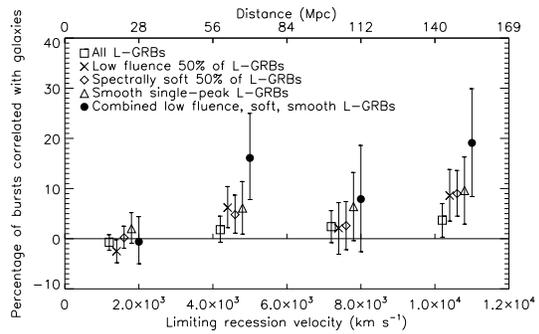, width=3.0in}
\centering{\caption[Correlation of bursts versus concentric shells of galaxies in the \textit{PSCz} catalogue]{\label{PSCzshells}Measured correlation between BATSE long duration GRBs and galaxies in concentric shells of increasing galactic recession velocity from the \textit{PSCz} catalogue. Details as Figure~\ref{bothcorr}.}}
\end{figure}

Turning now to the sub-samples with properties similar to known local bursts, Figure~\ref{PSCzshells} shows correlations with concentric \textit{shells} (as opposed to \textit{spheres}) of galaxies with recession velocity radii of $0-2000,~2000-5000,~5000-8000$ and $8000-11000~\rm{km~s^{-1}}$. From this figure it can be seen that each of the low-fluence, spectrally-soft, and smooth light curve sub-samples exhibits broadly equivalent, marginally increased correlations in the second and fourth shells. Furthermore, a combined set containing only those bursts (177) common to all three individual sub-samples exhibited increased correlation of $16\%\pm8\%$ ($\equiv28\pm14$ bursts) in the $2000-5000~\rm{km~s^{-1}}$ shell, and $19\%\pm11\%$ ($\equiv34\pm19$ bursts) in the outermost shell. 

\begin{figure}
\psfig{figure=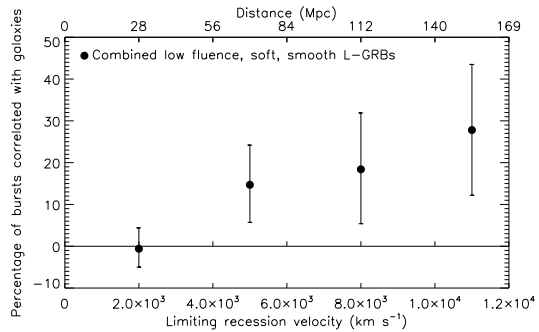, width=3.0in}
\centering{\caption[Correlation of low-fluence, spectrally soft, single-peaked L-GRBs versus concentric spheres of galaxies in the \textit{PSCz} catalogue]{\label{PSCzspheres}Measured correlation between the combined low-fluence, spectrally soft, single-peaked L-GRBs and galaxies in concentric spheres from the \textit{PSCz} catalogue.}}
\end{figure}

Returning to examine concentric \textit{spheres} of galaxies, Figure~\ref{PSCzspheres} shows the cumulative correlation versus radius of the combined sub-sample of L-GRBs, where it can be seen that $28\%\pm16\%$ of low-fluence, spectrally-soft bursts with smooth single-peaked light curves are correlated with galaxies within $\approx155~\rm{Mpc}$, equivalent to a total of $50\pm28$ bursts in the 9 years of BATSE operation.

It is interesting to ask whether our analysis is more sensitive to individual burst/galaxy correlations, or to correlation with large scale structure on the sky. In order to address this question, we repeated our correlation analyses but this time in calculation of $\Phi_{100}$, for each correlated pseudo-burst we removed the galaxy from which its position was generated. Thus $\Phi_{100}$ then measures the correlation between pseudo-bursts and the large scale structure around the generating host. The resulting structure-based $\Phi_{100}$ values for the \textit{PSCz} galaxy shells are shown in Table~\ref{phi100table}.

\begin{table}
{\centering
\begin{tabular}{ccc}
\hline
\textit{PSCz} galaxy shell&$\Phi_{100}$&$\Phi_{100}$\\
recession velocity&incl. host&excl. host\\
($\rm{km~s}^{-1}$)&(arbitrary units)&(arbitrary units) ($\%$)\\
\hline
$\le$2000&1875&1684 (90$\%$)\\
2000--5000&2873&2663 (93$\%$)\\
5000--8000&2354&2151 (91$\%$)\\
8000--11000&1810&1616 (89$\%$)\\
\hline
\end{tabular}
}
\caption{Values of $\Phi_{100}$ for \textit{PSCz} galaxy samples measured including and not including the host of the pseudo-bursts in correlated simulations}
\label{phi100table}
\end{table}
  
As this table shows, $\ge89\%$ of the correlation signal measured by $\Phi_{100}$ is generated from galaxies other than the specific hosts of the pseudo-bursts. It would therefore seem likely that correlation between the real BATSE bursts and galaxy distributions is mainly due to large scale structure on the sky rather than correlation with individual hosts. Given the size of the BATSE error boxes, this is certainly not surprising, but means that our analysis is unlikely to be sensitive to individual galaxy properties. Indeed, analyses performed weighting $\Phi$ by estimated individual galactic mass or star formation rate (estimated using the methods described in~\citet{2006MNRAS.368L...1L}), showed no significant change from the unweighted results. However, given the often large error boxes of individual bursts, sensitivity to structure is actually an advantage. Our technique effectively measures the correlation between two 2-dimensional signals, and is statistical in its nature - the signal is produced from the sums of contributions of many burst-galaxy candidate pairs and thus enhances any inherently weak individual correlation measurements at the expense of sensitivity to individual pair properties. This also precludes our ability to identify individual burst/galaxy associations. It is also known that the typical hosts of L-GRBs are blue, faint, irregular galaxies~\citep{2006Natur.441..463F} and thus many potential individual hosts may be missing from the galaxy catalogues.

\section{Discussion}

\subsection{Two populations of long bursts?}
\begin{figure}
\psfig{figure=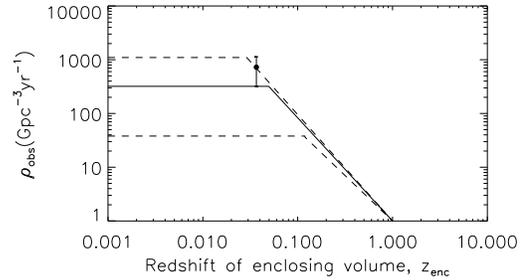, width=3.0in}
\centering{\caption[Observed burst rate versus redshift of enclosing volume]{\label{liang}The observed burst rate, $\rho_{\rm{obs}}$, within the enclosed volume out to redshift ${z}_{\rm{enc}}$. The filled circle shows the observed burst rate within our investigated total volume. The area enclosed by the dashed lines represents the upper and lower limits of the predicted observed rate for Low-Luminosity GRBs as a function of enclosing volume (after Figure 5(\textit{b}) of \citet{2007ApJ...662.1111L}).}}
\end{figure}

Our results confirm that nearby long GRBs as a whole are indeed rare events, with correlation remaining consistent with zero out to $\approx155~\rm{Mpc}$, but with a $10\%$ upper limit ($1\sigma$), equivalent to 144 bursts. Restricting the L-GRB sample to those with properties similar to known nearby bursts increased the measured correlation to $28\%\pm16\%$ of the sub-sample ($\equiv50\pm28$ bursts) within the same radius. It is worth emphasising that this means the 177 burst sub-sample contains a \textit{lower} limit ($1\sigma$) of 22 bursts correlated with local large scale structure, almost one quarter more than the $1\sigma$ \textit{upper} limit (18) expected from the correlation rate of the sample as a whole. The local rate density of under-luminous L-GRBs implied by our result is in agreement with those calculated via detector sensitivity and luminosity function arguments. For example,\citet{2006Natur.442.1014S} argue that sub-energetic bursts are 10 times more abundant than typical bright GRBs based on the sensitivities of \textit{BeppoSAX}, \textit{HETE-2} and \textit{Swift} to GRB\,980425 and GRB\,060218, and similarly ~\citet{2006Natur.442.1011P} (using BATSE, \textit{HETE-2} and \textit{Swift} sensitivities) found a local rate density at least 100 times greater than that estimated from cosmological bursts alone. In addition~\citet{2006ApJ...645L.113C} estimate an event ratio $\sim10^2$ between low and high luminosity bursts based on assuming the \textit{Swift} population of high-luminosity bursts to be complete to its mean redshift. Furthermore, \citet{2007ApJ...662.1111L} suggest that in order to avoid over-predicting the number of intermediate luminosity GRBs observed at low redshifts, the high-luminosity and low-luminosity bursts must be characterised by separate luminosity functions. They choose models of the same functional form (smoothed broken power-law), but with the coefficients of each separately constrained to produce GRB rates consistent with those observed. Finally, \citet{2007ApJ...657L..73G} estimate a local GRB rate of $200-1800\rm{Gpc}^{-3}\rm{yr}^{-1}$ using a luminosity function consistent with the luminosities of local bursts. Figure~\ref{liang} shows our results plotted with respect to predicted rates of Low-Luminosity GRBs from \citet{2007ApJ...662.1111L}. The results presented here can be seen to provide independent confirmation of these luminosity function based predictions. 

\subsection{Comparison with supernova searches}

GRB\,980425 was the first GRB to be observationally associated with a supernova~\citep{1998Natur.395..670G}, and GRB\,060218~\citep{2006GCN..4775....1C} is the most striking recent example. These nearby SN associated L-GRBs share similar properties (low-fluence, spectrally soft and smooth single-peaked light curves) as argued by \citet{1998ApJ...506L.105B}, and it may therefore seem that a search for temporal and spatial correlation between bursts and supernovae would provide a means of identifying individual GRB hosts, particularly at low redshift. However the heterogeneity and inherent incompleteness of SN catalogues (due for example to the difficulty of SN detection in dusty environments, biased surveys and magnitude limitations) make it difficult to use them to identify GRB host galaxies. Recent observations also suggest that not all long GRBs are necessarily accompanied by an observable supernova~\citep{2006Natur.444.1047F,2006Natur.444.1053G,2006Natur.444.1050D}, and of course, not all Ib/c SNe produce GRBs (e.g.~\citet{2006Natur.442.1014S}). Nevertheless several groups have attempted searches for spatial and temporal coincidences of GRBs with SNe (for example \citet{1998ApJ...504L..87W,1998ApJ...506L..27K,1999ApJ...518..901N}). Results obtained vary depending particularly on any restrictions imposed on the burst or SN properties in the analyses, and generally the statistics remain too poor to draw any firm conclusions from these and other studies. In addition, it has been suggested recently that the host galaxies of SNe type Ic without GRBs may be systematically different from those with GRBs, particularly in terms of metallicity~\citep{2007astro.ph..1246M}.

It is therefore apparent that using SNe searches to identify possible nearby GRBs is fraught with difficulty. \citet{2006Natur.442.1014S} estimate that only $\sim3\%$ of Ib/c SNe give rise to detectable low-luminosity GRBs comparable to GRBs\,980425 and 060218.  During the lifetime of the BATSE experiment, the Asiago supernova catalogue~\citep{1999A&AS..139..531B}\footnote{http://web.pd.astro.it/supern/snean.txt} contains only 33 events classified as Ib, Ibc or Ic within the volume considered here. Given further that most supernova searches, particularly in the 1990s, concentrated on targetting relatively bright galaxies, untypical of GRB hosts, we should expect that few, if any, GRB-SN were discovered by chance in these surveys. Our approach, independent of any assumed correlation other than that GRBs must occur in or near a host galaxy, is therefore justified and enables the placing of limits to the number of observed nearby L-GRBs based on location with respect to potential host galaxy distributions alone.

\section{Summary and Conclusions}

We have analysed the correlations of the entire BATSE catalogue of long GRBs with measured fluences localised to better than $10^{\circ}$ (1437 bursts) with galaxy samples out to $\approx155~\rm{Mpc}$ from two independent galaxy catalogues. We find that correlation between the L-GRB set as a whole and samples from both galaxy catalogues remains within 1 standard deviation ($\sigma$) of zero out to the highest radii considered. However, selecting a sub-sample of bursts with properties similar to those of the known local L-GRBs significantly increased correlation with large scale structure on the sky to a level of $28\%\pm16\%$ ($\equiv50\pm28$ bursts) within the same radius.

The cumulative correlation rate out to $\approx155~\rm{Mpc}$ suggests that BATSE most likely observed between 2 and 9 long GRBs per year similar to, and from within similar distances to the closest known GRBs to date. This implies an observed local L-GRB rate density within 155 Mpc of $700\pm360~\rm{Gpc}^{-3}\rm{yr}^{-1}$ (to BATSE detection limits, and assuming the BATSE sky exposure fraction to be of the order of 0.5). This is in reasonable agreement with $230^{+490}_{-190}~\rm{Gpc}^{-3}\rm{yr}^{-1}$ as calculated by \citet{2006Natur.442.1014S} via combined analyses of the sensitivities of \textit{BeppoSAX}, \textit{HETE II}, \textit{Swift}. It is also in very good agreement with observed rates of Low-Luminosity GRBs predicted from luminosity function arguments~\citep{2007ApJ...662.1111L,2007ApJ...657L..73G}.

At the time of writing (March 2007) $\approx50$ \textit{Swift}-detected long duration GRBs have measured redshifts out of a total of $\approx180$ detections. The mean redshift of an unbiased sample of 16 \textit{Swift} long GRBs measured in late 2005 was $z=2.8$~\citep{2006A&A...447..897J}, though the highest so far is  $z=6.295$~\citep{2006Natur.440..181H,2006Natur.440..184K} and the lowest GRB\,060218 with $z=0.0331$. Comparison between datasets from different missions and detectors is difficult (see for example~\citet{2005ApJ...634..501B,2006ApJ...644..378B}). However if we naively assume the \textit{Swift} GRB catalogue to be similar to the BATSE catalogue, then in a sample of 50 bursts we would expect to find between 0 and 5 within 155Mpc, as is the case. It may be argued that since we are looking for nearby bursts, these would be more likely to have measured redshifts than the bursts in general, though all three nearest GRBs were significantly underluminous events. Even so, comparing to the entire \textit{Swift} sample of 180 long GRBs, our results suggest it contains $9\pm9$ bursts with redshift less than 0.0367. So far, only one has been identified as occurring within our considered volume: GRB\,060218. This is certainly consistent with our limits (particularly given the relatively small numbers involved), but we may expect more low-redshift long bursts to be identified in the future, particularly among spectrally-soft, under-luminous bursts with little variability.

\section{Acknowledgements}
We thank Pall Jakobsson for useful discussions. RC and RSP acknowledge the support of the University of Hertfordshire. NRT and AJL acknowledge the support of UK PPARC senior and postdoctoral research fellowships respectively.

\bibliographystyle{mn2e_281004}
\bibliography{bobrefs}

\label{lastpage}

\end{document}